\documentclass[conference]{IEEEtran}
\usepackage{fancyhdr} 
\ifCLASSINFOpdf
  \usepackage[pdftex]{graphicx}

\else

\fi

\usepackage{amsmath}
\usepackage{amssymb}

\usepackage{algorithmic}

\usepackage{array}

\IEEEoverridecommandlockouts
\usepackage{cite}
\usepackage{amsfonts}
\usepackage{textcomp}
\usepackage{xcolor}
\def\BibTeX{{\rm B\kern-.05em{\sc i\kern-.025em b}\kern-.08em
    T\kern-.1667em\lower.7ex\hbox{E}\kern-.125emX}}

\begin{document}
%
\title{Graph Neural Network Based Hybrid Beamforming Design in Wideband Terahertz MIMO-OFDM Systems
\thanks{This work was supported in part by the National Science Foundation under ECCS Grant 2234122.}
}

\author{\IEEEauthorblockN{Beier Li}
\IEEEauthorblockA{Tufts University\\
Medford, MA 02155\\
Beier.Li@tufts.edu}
\and
\IEEEauthorblockN{Mai Vu}
\IEEEauthorblockA{
Tufts University\\
Medford, MA 02155\\
Mai.Vu@tufts.edu}}


%


\maketitle

\IEEEpubid{\begin{minipage}{\textwidth}\ \\[60pt]
 \hspace*{4em} 979-8-3503-9214-2/24/\$31.00 \copyright 2024 IEEE
\end{minipage}}

\IEEEpubidadjcol

\begin{abstract}
6G wireless technology is projected to adopt higher and wider frequency bands, enabled by highly directional beamforming. However, the vast bandwidths available also make the impact of beam squint in massive multiple input and multiple output (MIMO) systems non-negligible. Traditional approaches such as adding a true-time-delay line (TTD) on each antenna are costly due to the massive antenna arrays required. This paper puts forth a signal processing alternative, specifically adapted to the multicarrier structure of OFDM systems, through an innovative application of Graph Neural Networks (GNNs) to optimize hybrid beamforming. By integrating two types of graph nodes to represent the analog and the digital beamforming matrices efficiently, our approach not only reduces the computational and memory burdens but also achieves high spectral efficiency performance, approaching that of all digital beamforming. The GNN runtime and memory requirement are at a fraction of the processing time and resource consumption of traditional signal processing methods, hence enabling real-time adaptation of hybrid beamforming. Furthermore, the proposed GNN exhibits strong resiliency to beam squinting, achieving almost constant spectral efficiency even as the system bandwidth increases at higher carrier frequencies.
\end{abstract}


%
\IEEEpeerreviewmaketitle

\section{Introduction}
Hybrid beamforming, which combines analog and digital techniques, offers a cost-effective solution and robust performance for employing massive antenna arrays in modern communication systems. In wideband systems utilizing the OFDM technique to enhance data rates and resistance to multipath effects, the beam pattern increasingly varies with frequency changes across different subcarriers \cite{beamsquint}. This phenomenon, known as beam squint, becomes particularly significant in 6G wireless networks operating in the sub-Terahertz spectrum from 100 GHz to 1 THz, where the bandwidth is much wider than 5G systems, of around 18 GHz \cite{THz}.

To manage beam squint in wideband systems with hybrid beamforming, recent research efforts have focused on two main solutions: true-time-delay lines (TTD) and signal processing methods. TTD is a time-delay filter integrated into each antenna to provide precise control over signal timing and effectively resolve beam squint problems \cite{TTD_subarray, TTD_hybridfield}. However, in the Terahertz frequency range, the ability to pack more antennas into the same device size not only enhances performance but also substantially increases the number of TTDs required, leading to substantially higher costs. Given these economic considerations, signal processing methods become an attractive alternative for managing beam squinting in high-frequency domains because of their cost-efficiency.

Despite progress in the field, current literature indicates that beamforming designs for wideband channels lack efficient solutions to effectively address the beam squint issue \cite{sparse, sixmethods}. Algorithms such as Alternative Manifold Optimization (AMO) \cite{AMO} and Iterative Coordinate Descent (ICD) \cite{ICD} have been proposed, but they suffer from the need for continuous optimization with every channel update and hence require significant computational resources, limiting their practical advantage.

In contrast, machine learning offers a promising alternative that can significantly reduce the computational burden. Specifically, the unique structure of OFDM systems has inspired us to construct a graph with two different types of nodes that can effectively train a model through Graph Neural Networks (GNN). By doing so, we not only can optimize beamforming design more efficiently but also achieve performance that closely matches that of traditional numerical optimization algorithms, providing a new practical pathway to address the beam squint problem in wideband channels.

\section{System Model And Problem Formulation}
\subsection{System Model}
\begin{figure*}[!t]
\centering
\includegraphics[width=6in]{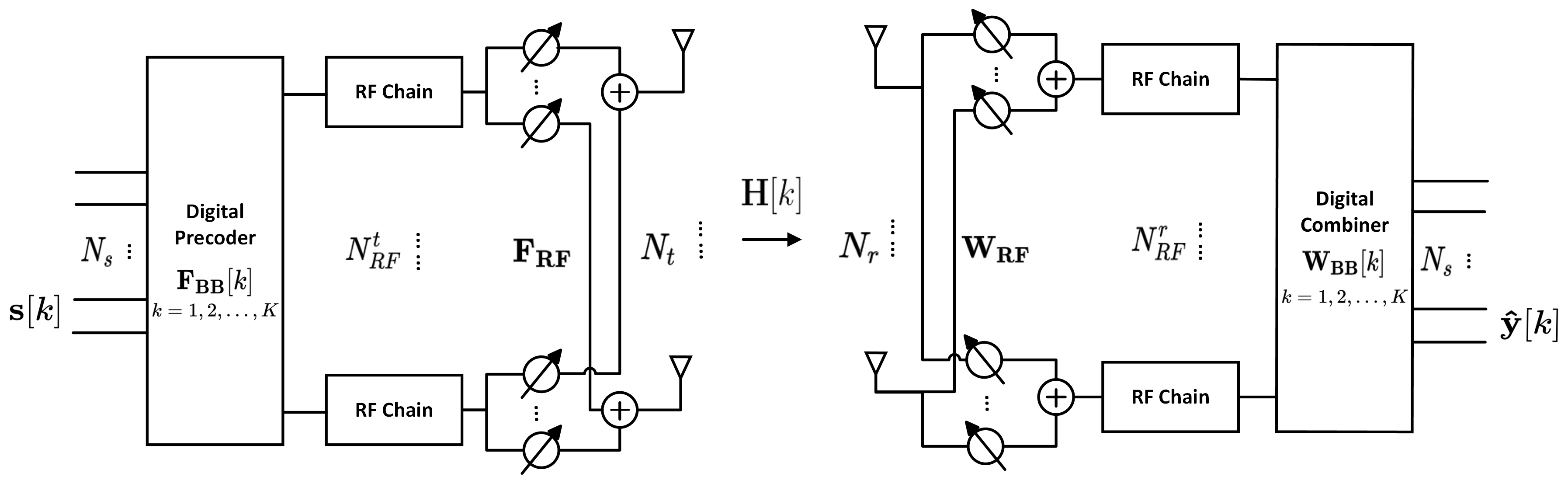}
\caption{Block diagram of a single-user MIMO-OFDM system with hybrid beamforming architecture at the BS and the UE.}
\label{fig_sys}
\end{figure*}

We consider a single-user MIMO-OFDM system as shown in Fig. \ref{fig_sys}, where $N_s$ data streams are sent by the base station (BS), which is equipped with $N_{\text{RF}}^t$ RF chains and $N_t$ antennas. At the receiver, we have $N_r$ antennas and $N_{\text{RF}}^r$ RF chains. Further, it holds that $N_s \leq N_{\text{RF}}^t \leq N_t$ and $N_s \leq N_{\text{RF}}^r \leq N_r$ because of hardware constraint. 

The transmitted signal at the $k$-th subcarrier can be written as $\mathbf{x}[k] = \mathbf{F_{\text{RF}}F_{BB}}[k]\mathbf{s}[k]$, where $\mathbf{s}[k]$ is the $N_s \times 1$ symbol vector conveyed by each subcarrier $k = 1, 2, ..., K$, and assumed that $\mathbb{E}[\mathbf{s}[k]\mathbf{s}^*[k]]=\mathbf{I}_{N_s}$. The hybrid beamformers consist of a digital baseband beamformer $\mathbf{F_{BB}}[k] \in \mathbb{C}^{N_{\text{RF}}^t \times N_s}$ and the analog RF beamformer $\mathbf{F_{\text{RF}}} \in \mathbb{C}^{N_t \times N_{\text{RF}}^t}$. The normalized transmit power constraint is given by $||\mathbf{F_{\text{RF}}F_{BB}}[k]||_F^2=1$. Thus, the received signal at the $k$-th subcarrier is
\begin{align}
    \mathbf{\hat{y}}[k] =& \sqrt{P_r}\mathbf{W_{BB}^*}[k]\mathbf{W_{\text{RF}}^*}\mathbf{H}[k]\mathbf{F_{\text{RF}}F_{BB}}[k]\mathbf{s}[k]\nonumber\\&+\mathbf{W_{BB}^*}[k]\mathbf{W_{\text{RF}}^*}\mathbf{n}[k],
\end{align}
where $\mathbf{H}[k]$ is the channel matrix at the $k$-th subcarrier, and $\mathbf{n}[k]\sim\mathcal{CN}(0,\sigma_n^2\mathbf{I}_{N_r})$ is the additive white Gaussian noise, with the digital baseband combiner $\mathbf{W_{BB}}[k] \in \mathbb{C}^{N_{\text{RF}}^r \times N_s}$ and analog RF combiner $\mathbf{W_{\text{RF}}} \in \mathbb{C}^{N_r \times N_{\text{RF}}^r}$. $P_r$ stands for the average received power.

In this paper, we assume that the perfect channel state information (CSI) is known, which can be obtained by channel estimation \cite{CSI}. Then the achievable spectral efficiency can be expressed as
\begin{align}
    R =& \frac{1}{K}\sum_{k=1}^K \log_2\left[\det\left(\mathbf{I}_{N_s} + \frac{P_r}{\sigma_n^2}\mathbf{W_{BB}^*}[k]\mathbf{W_{\text{RF}}^*}\mathbf{H}[k]\mathbf{F_{\text{RF}}F_{BB}}[k] \right.\right.\nonumber \\
    &\quad \biggl.\biggl. \times \mathbf{F_{BB}^*}[k]\mathbf{F_{\text{RF}}^*H^*}[k]\mathbf{W_{\text{RF}}}\mathbf{W_{BB}}[k]\biggr)\biggr]
\end{align}

\subsection{Channel Model}
We adopted a clustered double-directional small-scale channel model \cite{channel}:
\begin{align}
    \mathbf{H}[k] = \sqrt{\frac{N_tN_r}{N_{\text{cl}}N_{\text{ray}}}} \sum_{i=1}^{N_{\text{cl}}} \sum_{l=1}^{N_{\text{ray}}} \alpha_{il} \mathbf{a}_r(\phi_{il}^r, \theta_{il}^r) \mathbf{a}_t(\phi_{il}^t, \theta_{il}^t)^*,
\end{align}
such that $\mathbb{E}[||\mathbf{H}[k]||_F^2]=N_tN_r$. Here $N_{\text{cl}}$ and $N_{\text{ray}}$ represent the number of clusters and the number of rays in each cluster, and $\alpha_{il}$ denotes the complex gain of the $l$-th ray in the $i$-th cluster, where $\alpha_{il}$ follows the distribution $\mathcal{CN}(0,\sigma_{i}^2)$. $(\phi^r_{il}, \theta^r_{il})$ and $(\phi^t_{il}, \theta^t_{il})$ are the azimuth and elevation angles of arrival and departure, respectively. Considering the uniform planar array (UPA) antenna elements, the array response vector corresponding to the $l$-th ray in the $i$-th cluster can be written as
\begin{align}
    \mathbf{a}(\phi_{il}, \theta_{il}) = &\frac{1}{\sqrt{MN}} \left[ 1, ..., e^{j\frac{2\pi}{\lambda_k}d(p\sin\phi_{il}\sin\theta_{il}+q\cos\theta_{il})},...,\right.\nonumber\\
    &\left.e^{j\frac{2\pi}{\lambda_k}d((M-1)\sin\phi_{il}\sin\theta_{il}+(N-1)\cos\theta_{il})} \right]^T,
\end{align}
where $d$ and $\lambda$ are the antenna spacing and the signal wavelength, $p$ and $q$ are the antenna indices in the 2D plane.

\subsection{Problem Formulation}
As shown in \cite{sparse}, the problem can be separated into two subproblems to deal with the transmit beamformers and receive combiners separately. Since they have similar mathematical formulations, we will focus on the beamformer design in this paper. The proposed problem formulation is given by:
\begin{align}
    \max_{\mathbf{F_{\text{RF}}, F_{BB}}[k]} \quad & \frac{1}{K}\sum_{k=1}^K \log_2[\det(\mathbf{I}_{N_r} + \frac{P_t}{\sigma_n^2}(\mathbf{H}[k]\mathbf{F_{\text{RF}}F_{BB}}[k]\nonumber\\
    &\times \mathbf{F_{BB}^*}[k]\mathbf{F_{\text{RF}}^*H^*}[k]))]\nonumber\\
	\text{s.t.} \quad & |[\mathbf{F_{\text{RF}}}]_{i,j}|^2=1, \quad \forall i,j\nonumber\\
			&||\mathbf{F_{\text{RF}}F_{BB}}[k]||_F^2=1,\label{prob}
\end{align}
where $P_t$ stands for the average transmit power, with a constant modulus constraint on the analog beamforming component, which is introduced by the hardware constraint of the phase shifters, and a normalized transmit power constraint.

\section{Proposed Graph Neural Network Model}
In addressing the problem as shown in (\ref{prob}), the only known parameter is the channel model $\mathbf{H} \in \mathbb{C}^{N_r \times N_t \times K}$. In conventional deep neural networks (DNN), directly using the channel model as an input can lead to substantial storage complexity due to the massive antenna array and numerous subcarriers. However, we observe that a common analog beamformer is shared among all K sub-carriers corresponding to $K$ distinct digital beamformers. This insight has led us to employ an efficient Graph Neural Network (GNN) architecture having a bipartite structure with one node representing the analog beamformer and $K$ other nodes representing the digital beamformers in $K$ subcarriers. We adopt a message-passing mechanism within the GNN, utilizing the channel information in each subcarrier as the graph feature to determine the node representations for both analog node and digital nodes. These representations are then utilized to reconstruct the beamforming matrices.

\subsection{Graph Representation}
\begin{figure}[!t]
\centering
\includegraphics[width=3in]{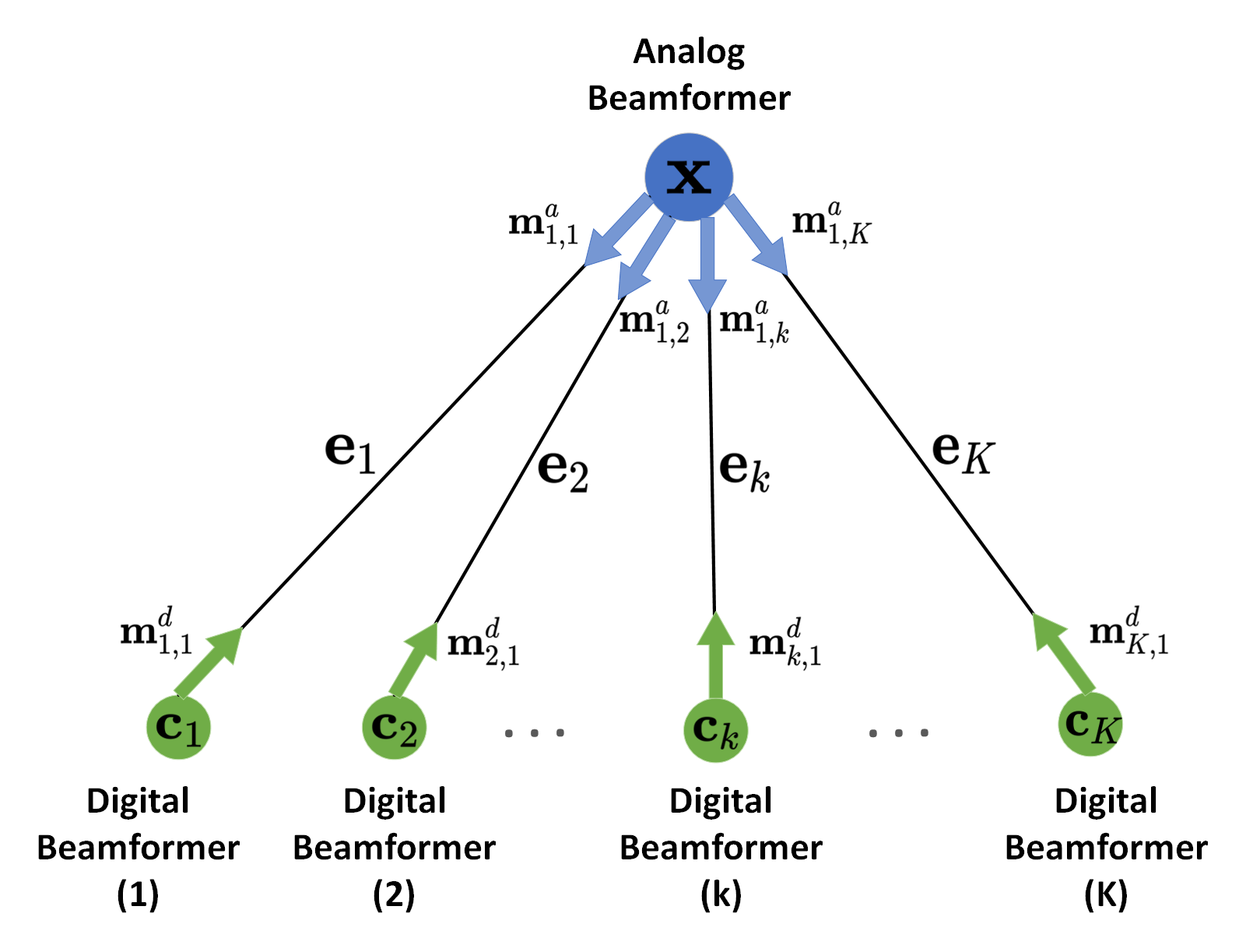}
\caption{Graph representation of a hybrid beamforming structure with two types of nodes representing the analog and the digital beamformers, where each edge has an edge feature $\mathbf{e}_k$. Here we implement a message-passing mechanism where messages are denoted as $\mathbf{m}_{1,k}^a$ and $\mathbf{m}_{k,1}^d$.}
\label{fig_graph}
\end{figure}

We construct an undirected graph $\mathcal{G}=(\mathcal{V},\mathcal{E})$ where $\mathcal{V}$ and $\mathcal{E}$ represent the sets of nodes and edges, respectively. In this graph, the analog beamformer is represented as an analog node, and the digital beamformers for different subcarriers are represented as $K$ digital nodes. These nodes form a fully connected bipartite graph, with an edge feature as $\mathbf{e}_k$ for each subcarrier $k = 1, ..., K$. These features, represented by the vectorized channel state information, serve as both the edge features and known information. Since the wireless channel coefficients are complex, we separate the real and imaginary parts and concatenate them into a real-valued vector as the input of the GNN. The edge feature can be expressed as
\begin{align}
\mathbf{e}_k=\left[ \text{vec} \left( \text{Re} \left\{ \mathbf{H}[k] \right\} \right)^T, \text{vec} \left( \text{Im} \left\{ \mathbf{H}[k] \right\} \right)^T \right]^T.\label{edge}
\end{align}
The graph is described in Fig. \ref{fig_graph}.
As shown in the graph, we adopt a message-passing mechanism that collects both node and edge information from each node's neighbors to form messages $\mathbf{m}^a_{1,k}$ for the analog node and $\mathbf{m}^d_{k,1}$ for the digital nodes along their respective edges. This information is then aggregated to update the node representations, effectively capturing the unique structure of this bipartite graph. Our goal is to update every node representation vector in the graph and train the GNN model to achieve a high average data rate. Ultimately, the updated node representations are reconstructed into the desired beamforming matrices.

\subsection{GNN Structure}
\begin{figure*}[!t]
\centering
\includegraphics[width=5.2in]{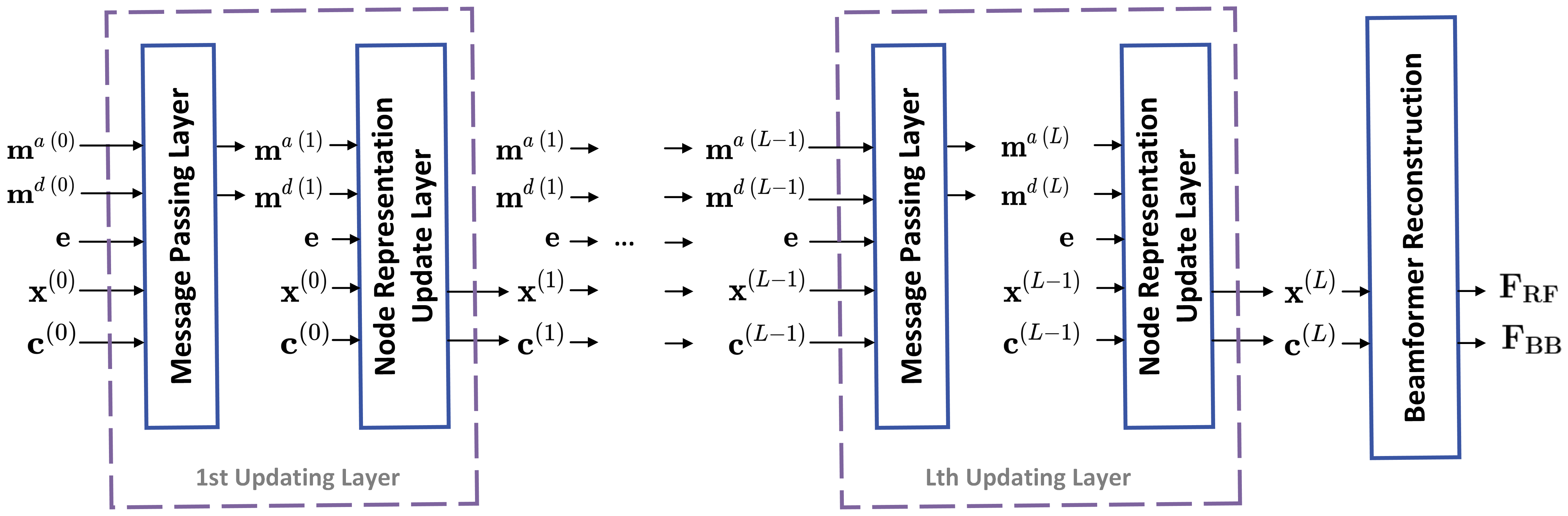}
\caption{The overall architecture of the proposed GNN consists of $L$ updating layers, each integrating a message-passing layer and a node representation updating layer, followed by a final beamformer reconstruction layer.}
\label{fig_gnn}
\end{figure*}

The proposed GNN consists of $L$ updating layers,  each consisting of a message-passing layer and a node representation update layer, followed by a final beamformer reconstruction layer. The architecture is illustrated in Fig. \ref{fig_gnn}.

\subsubsection{Message Passing Layer}

At the $l$-th layer of the GNN, the messages sent from digital node $k$ to analog node $1$, and from analog node $1$ to digital node $k$ can be generated and described as
\begin{align}
&\mathbf{m}_{1,k}^{a\ (l)}=f_1^a(\mathbf{e}_k,\mathbf{x}^{(l-1)},\phi({\mathbf{m}_{k,1}^{d\ (l-1)}})_{k\in\mathcal{N}(1)}, \mathbf{m}_{1,k}^{a\ (l-1)})\\
&\mathbf{m}_{k,1}^{d\ (l)}=f_1^d(\mathbf{e}_k,\mathbf{c}_k^{(l-1)},{\mathbf{m}_{1,k}^{a\ (l-1)}}, {\mathbf{m}_{k,1}^{d\ (l-1)}})
\end{align}
where $\textit{f}_1^{\,a}(\cdot)$ and $\textit{f}_1^{\,d}(\cdot)$ are fully connected neural networks for the analog node and digital nodes, respectively. $\mathbf{e}_{k}$ is the edge feature described in (\ref{edge}). $\phi(\cdot)$ is the element-wise mean function to aggregate the messages, generated from the last layer, from the neighboring nodes. $\mathcal N(1)$ represents all the nodes that are connected to analog node $1$. By doing this aggregation operation, we can include the graph structure information in the generated messages by emphasizing that each digital beamformer and channel for every subcarrier is dedicated to a single analog beamformer.

\subsubsection{Node Representation Updating Layer} 

After generating the messages, each node receives them via the connecting edges, aggregates these messages from its neighbors, and then updates its representation vector accordingly as follows.
\begin{align}
&\mathbf{x}^{(l)}=f_2^a(\mathbf{x}^{(l-1)}, \phi({\mathbf{m}_{k,1}^{d\ (l)}})_{k\in\mathcal{N}(1)}, \phi(\mathbf{e}_k)_{k \in \mathcal{K}})\\
&\mathbf{c}_k^{(l)}=f_2^d(\mathbf{c}_k^{(l-1)}, \mathbf{m}_{1,k}^{a\ (l)}, \mathbf{e}_k)
\end{align}
$\textit{f}_2^{\,a}(\cdot)$ and $\textit{f}_2^{\,d}(\cdot)$ are two other fully connected neural networks for the analog node and digital nodes, respectively. Again, $\phi(\cdot)$ is the element-wise mean function to aggregate the messages. By utilizing this operation, we further emphasize the graph's structure and incorporate channel information to enhance the network data, thereby more effectively capturing the mathematical relationship between the channels and the beamforming matrices.

\subsubsection{Beamformer Reconstruction Layer} 

If we directly reconstruct our beamforming matrices using the node representation $\mathbf{x}$ and $\mathbf{c}_k$ from the $l$-th layer output of the network through reshaping, this approach would not satisfy the two constraints as shown in our initial problem (\ref{prob}). Therefore, we need to process them in the final layer of this network. The output analog node representation $\mathbf{x}^{(L)}$ is reshaped into the phase matrix $\mathbf{X}$ of the $\mathbf{F_{\text{RF}}}$ matrix to meet the constant modulus constraint. Meanwhile, the complex matrix $\mathbf{C}_k$ is reconstructed using $\mathbf{c}_k^{(L)}$ through normalization to fulfill the normalized power constraint. The analog and digital beamformers can be obtained as follows:
\begin{equation}
\mathbf{F_{\text{RF}}}= e^{j\mathbf{X}}, \quad
\mathbf{F_{BB}}[k] = \frac{\mathbf{C}_k}{|| \mathbf{C}_k e^{j\mathbf{X}}||_F^2}
\end{equation}
where
\begin{align}
    \mathbf{X}=&\text{reshape}(\mathbf{x}^{(L)},(N_t,N_{\text{RF}}^t)),\\
    \mathbf{C}_k=&\text{reshape}(\mathbf{c}_k^{(L)}[1:N_{\text{RF}}^t\times N_s],(N_{\text{RF}}^t,N_s)) \nonumber\\
    &+ j\ \text{reshape}(\mathbf{c}_k^{(L)}[N_{\text{RF}}^t\times N_s:2N_{\text{RF}}^t\times N_s],(N_{\text{RF}}^t,N_s)).
\end{align}
\subsection{Training Process}
During the offline training phase, we optimize all parameters of the GNN, denoted as $\mathbf{\Omega}$ for message generation and representation vector update together, to minimize the loss function below, which is directly formulated based on the objective function in (\ref{prob}):
\begin{align}
L(\mathbf{\Omega}) =& -\frac{1}{K}\sum_{k=1}^K \log_2[\det(I_{N_r} + \frac{P}{\sigma_n^2}(\mathbf{H}[k]\mathbf{F_{\text{RF}}F_{BB}}[k]\nonumber\\
&\times \mathbf{F_{BB}^*}[k]\mathbf{F_{\text{RF}}^*H^*}[k]))],
\label{loss}
\end{align}
which is computed using the network outputs along with the available channel information.

To minimize the loss (\ref{loss}), we adopt a mini-batch stochastic gradient descent (SGD) approach, which updates the parameters according to the following formula:
\begin{equation}
\mathbf{\Omega}^{(i+1)} \leftarrow \mathbf{\Omega}^{(i)} - \eta\nabla_{\mathbf{\Omega}}\mathbb{E}_{\mathcal{B}}\left[L(\mathbf{\Omega})\right],
\end{equation}
where $\eta$ represents the learning rate, and $\mathcal{B}$ denotes the mini-batch set.

After the training phase, we will have four trained networks: two for the analog node, $\textit{f}_1^{\,a}(\cdot)$and $\textit{f}_2^{\,a}(\cdot)$, and two shared among all digital nodes across different subcarriers, $\textit{f}_1^{\,d}(\cdot)$ and $\textit{f}_2^{\,d}(\cdot)$. It is important to note that since each subcarrier uses the same neural network, there is no need to train with a large number of subcarriers. Instead, we only need to ensure sufficient channel state information is covered across various frequencies. Therefore, we can choose a number of subcarriers, $K$, for training such as 4 or 8. Then during the online running phase, the number of digital nodes can be increased as needed to suit different OFDM systems.

\section{Numerical Simulations}

\subsection{Simulation System Settings}
In this section, simulation results are presented to show the performance of the proposed GNN structure. We use a carrier frequency of 300GHz and a bandwidth of 30GHz, with 8 subcarriers selected for the offline training process. The BS employs an $N_t = 64$ UPA antenna system, equipped with $N_s=N_{\text{RF}}^t=4$ RF chains, while the receiver uses a $N_r = 8$ UPA antenna system. The channel parameters are defined with $N_{\text{cl}}=2$ clusters and $N_{\text{ray}}=2$ rays per cluster. Each cluster's average power is set to $\sigma_{i}^2 = 1$. Both the azimuth and elevation angles of departure and arrival (AoDs and AoAs) are modeled to follow a Laplacian distribution, with uniformly distributed mean angles and an angular spread of 10 degrees \cite{THz, channel}. The antenna elements are spaced at half the wavelength.

\subsection{Offline GNN Training}
\begin{figure}[!t]
\centering
\includegraphics[width=3.5in]{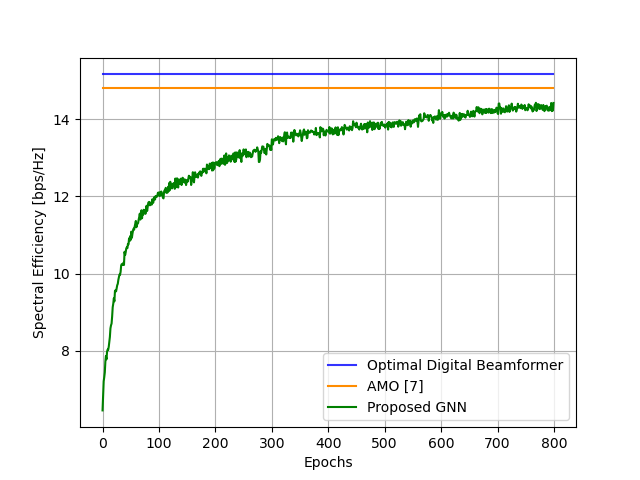}
\caption{Convergence of the proposed GNN training for SNR = 0dB, $N_t=64$, $N_{\text{RF}}^t=4$, $K=8$, 
 $f_c=300$GHz, and $B=30$GHz.}
\label{fig_train}
\end{figure}

For training, the initialization of the analog node representation follows a uniform distribution over $[0,2\pi)$, while the initial value of the digital node representations follows a Gaussian distribution. The neural networks implemented at each node consist of two hidden layers, each containing twice as many neurons as the input size. We employ the Adam optimizer with a learning rate of $2\times 10^{-4}$ for training. During the training process, we observed performance improvements by reducing the learning rate by half every 300 epochs, which is adopted as a strategy that effectively enhances the model performance. For training, each mini-batch consists of 100 samples, and we process 100 such batches per epoch to update the parameters of the neural networks. The GNN model is structured with $L=2$ layers. Fig. \ref{fig_train} shows the convergence of the proposed GNN structure, compared with the optimal fully digital beamformer case, and the iterative optimization AMO algorithm in \cite{AMO}. We can see that the GNN model converges to a network spectral efficiency approaching that of the traditional optimization in \cite{AMO} and not far from fully digital beamforming performance.

\begin{figure}[!t]
\centering
\includegraphics[width=3.5in]{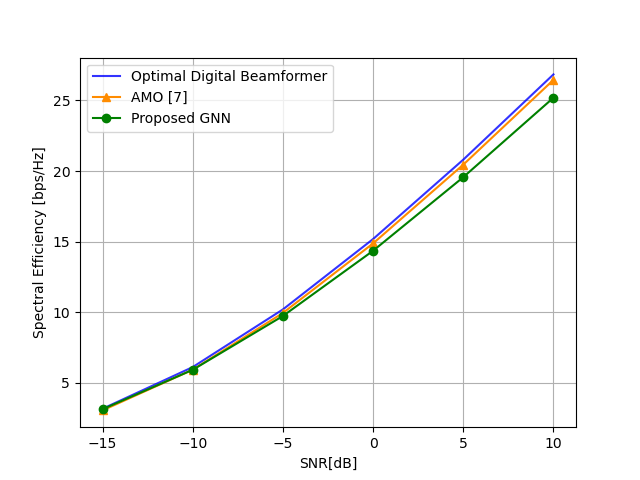}
\caption{Spectral efficiency achieved by different beamforming design algorithms with $K=64$ subcarriers, averaged over $10^3$ channel realizations.}
\label{fig_run}
\end{figure}

\subsection{Online GNN Inference}
During the online running phase, we increased the number of subcarriers to $K=64$ for simulation purposes, and all presented simulation results are averaged over $10^3$ channel realizations. As shown in Fig. \ref{fig_run}, as the SNR decreases, our proposed GNN closely approximates the performance of the optimal fully digital case. This observation suggests that our model is robust at a low SNR range. As the SNR increases, the performance gap between our model and the numerical optimization algorithm remains within an acceptable range. This indicates that our GNN model maintains competitive effectiveness at higher SNR levels, offering a practical alternative to traditional numerical methods.

\begin{figure}[!t]
\centering
\includegraphics[width=3.5in]{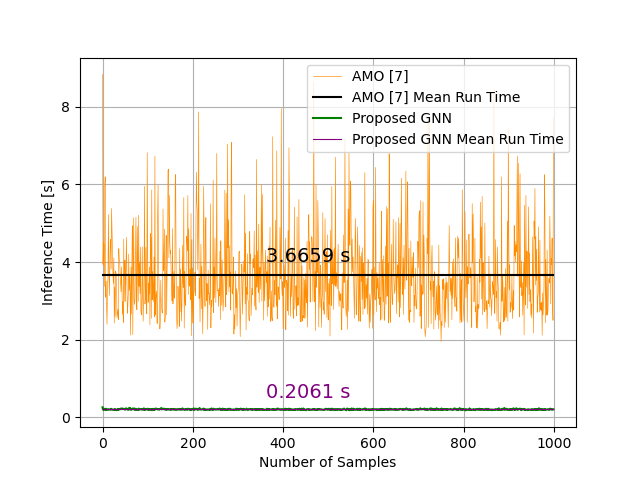}
\caption{Runtime comparison per CSI update on NVIDIA V100 GPU, for $10^3$ CSI samples.}
\label{fig_time}
\end{figure}

The practicality of the GNN can be seen more clearly in the runtime comparison in Table \ref{time} and Fig. \ref{fig_time}. Here, we compare the running times for the two algorithms across all $10^3$ channel realizations. It is clear that the AMO algorithm requires significantly more time, around 18 times or more than an order of magnitude slower, than the proposed GNN algorithm. This difference is due to the AMO algorithm needing to repeat the alternative optimization process for each subcarrier and each channel realization, while the GNN simply utilizes the pre-trained model to perform feed-forward computation, producing results directly by scaling up the number of digital nodes. Furthermore, the variance in computation time of the GNN is significantly smaller, at two orders of magnitude smaller, than of the AMO, giving the GNN a very stable computation time per CSI update. Both the average run time and standard deviation demonstrate the GNN's suitability for real-time array steering in practice.

\begin{table}[!t]
\renewcommand{\arraystretch}{1.3}
\caption{Simulation Runtime Comparison Per CSI Update on NVIDIA V100 GPU, Averaged Over $10^3$ CSI Samples}
\label{time}
\centering
\begin{tabular}{|c||c|c|}
\hline
Method & AMO \cite{AMO} & Proposed GNN\\
\hline
Mean Run Time (sec) & 3.6659 & 0.2061\\
\hline
Standard Deviation (sec) & 1.0905 & 0.0107\\
\hline
\end{tabular}
\end{table}

\begin{table}[!t]
\renewcommand{\arraystretch}{1.3}
\caption{Simulated Dynamic Memory Allocation Comparison Per CSI Update on NVIDIA V100 GPU, Averaged Over $10^3$ CSI Samples}
\label{mem}
\centering
\begin{tabular}{|c||c|c|}
\hline
Method & AMO \cite{AMO} & Proposed GNN\\
\hline
Mean Dynamic Memory Allocation (Mb) & 4635.6 & 1268.9\\
\hline
Standard Deviation (Mb) & 1850.5 & 1.5 $\times 10^{-5}$\\
\hline
\end{tabular}
\end{table}

The dynamic memory allocation comparison as shown in Table \ref{mem} further emphasizes the GNN's computational efficiency. Since we use the pre-trained GNN model directly during the online running phase, the amount of dynamic memory required for each channel remains constant at around 1268.9Mb. In contrast, the AMO algorithm needs to perform repeated optimization for each channel realization, leading to significant variability in dynamic memory allocation, with a high standard deviation depending on the channel realization. Furthermore, the average memory usage of the AMO algorithm is almost 4 times higher than that of the GNN model. This analysis highlights the GNN's efficiency and stability in resource allocation, making it more suitable for practical deployment in large-scale antenna array systems while conserving computational resources.

\begin{figure}[!t]
\centering
\includegraphics[width=3.5in]{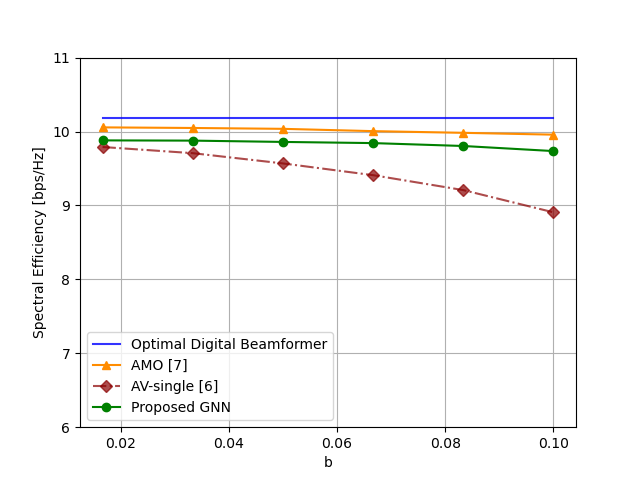}
\caption{Spectral efficiency versus channel bandwidth for different beamforming design algorithms, with $K=64$ subcarriers, SNR$=-5$dB, averaged over $10^3$ channel realizations. The central carrier frequency is $f_c = 300$GHz, and $b=\frac{B}{f_c}$, where $B$ is the communication channel bandwidth.}
\label{fig_beamsquint}
\end{figure}

Fig. \ref{fig_beamsquint} demonstrates the proposed GNN model's effective mitigation of the beamsquint issue. While comparing the spectral efficiency of AMO and the proposed GNN, we incorporated a new baseline algorithm in \cite{sixmethods}, which calculates the array response vector only for the central carrier frequency and uses that to directly construct $\mathbf{F_{RF}}$. This approach is directly affected by beam squinting, as it relies only on the central carrier frequency to construct the analog beamformer without considering the bandwidth or the number of subcarriers. Let $b=\frac{B}{f_c}$ represent the fractional bandwidth. A small value of $b$, close to 0, indicates negligible beam squinting. As $b$ increases, the beam squint effect becomes more pronounced as seen in the baseline AV-single performance \cite{sixmethods}. Our proposed GNN, on the other hand, effectively mitigates the beam squinting problem in wideband channels.

\section{Conclusion}
We proposed a novel GNN architecture for efficient hybrid beamforming design in wideband Terahertz OFDM-MIMO systems, while simultaneously mitigating the beam squint effect. By capturing the unique structure of hybrid beamforming in an OFDM system, we constructed a bipartite graph and utilized a message-passing mechanism to optimize the GNN performance. The proposed GNN model not only allows the optimization of both digital and analog beamforming matrices, but also adjusts them dynamically to changes in the number of subcarriers by scaling the digital nodes without the need for retraining the model. This method enhances both the system's spectral efficiency and adaptability in practical applications. Compared to traditional signal processing algorithms, our model offers significant competitive advantages in cost, running time, and memory resource requirements, making it viable for real-time beamforming adaptation.




%

\end{document}